\documentclass[10pt]{iopart}
\usepackage{graphicx}
\def\lesssim{\mathrel{\hbox{\rlap{\hbox{\lower4pt\hbox{$\sim$}}}\hbox{$<$}}}}
\def\gtrsim{\mathrel{\hbox{\rlap{\hbox{\lower4pt\hbox{$\sim$}}}\hbox{$>$}}}}
\def\MNRAS{{\it Mon.\ Not.\ Royal Astron.\ Soc. }}
\def\MNRASL{{\it Mon.\ Not.\ Royal Astron.\ Soc.\ Lett. }}
\def\ApJ{{\it Astroph.\ J. }}
\def\ApJL{{\it Astroph.\ J.\ Lett. }}
\def\ApJS{{\it Astroph.\ J.\ Supp. }}
\def\AA{{\it Astron.\ Astroph. }}
\def\AJ{{\it Astron.\ J. }}
\def\PRL{{\it Phys.\ Rev.\ Lett. }}
\def\PRD{{\it Phys.\ Rev.\ }D }
\def\CQG{{\it Clas.\ Quant.\ Grav. }}

\begin{document}

\title[Astrophysics of SMBH Mergers]{Astrophysics of Super-massive
  Black Hole Mergers}
\author{Jeremy D.\ Schnittman$^{1}$}
\address{$^1$ NASA Goddard Space Flight Center, Greenbelt, MD 20771}

\begin{abstract}
We present here an overview of recent work in the subject of
astrophysical manifestations of super-massive black hole (SMBH)
mergers. This 
is a field that has been traditionally driven by theoretical work, but
in recent years has also generated a great deal of interest and
excitement in the observational astronomy community. In particular,
the electromagnetic (EM) counterparts to SMBH mergers provide the
means to detect and characterize these highly energetic events at
cosmological distances, even in the absence of a space-based
gravitational-wave observatory. In addition to providing a mechanism
for observing SMBH mergers, EM counterparts also give important
information about the environments in which these remarkable events
take place, thus teaching us about the mechanisms through which
galaxies form and evolve symbiotically with their central black
holes. 
\end{abstract}

\submitto{\CQG}
\maketitle

%--------------------------------------------------------
\section{INTRODUCTION}
\label{intro}
%--------------------------------------------------------

Following numerical relativity's {\it annus mirabilis} of 2006, a
deluge of work has explored the astrophysical manifestations of black
hole mergers, from both the theoretical and observational
perspectives. While the field has traditionally been dominated by
applications to the direct detection of gravitational waves
(GWs), much of the recent focus of numerical simulations has been on
predicting potentially observable electromagnetic (EM) signatures. Of
course, the greatest science yield will come from coincident detection
of both the GW and EM signature, giving a myriad of observables such
as the black hole mass, spins, redshift, and host environment, all
with high precision \cite{bloom:09}. Yet even in the absence of a direct GW detection
(and this indeed is the likely state of affairs for at least the next
decade), the EM signal alone may be sufficiently strong to detect with
wide-field surveys, and also unique enough to identify unambiguously
as a SMBH merger.

In this article, we review the brief history and astrophysical
principles that govern the observable signatures of SMBH
mergers. To date, the field has largely been driven by theory, but we
also provide a summary of the observational techniques and surveys
that have been utilized, including recent claims
of potential detections of both SMBH binaries and also post-merger
recoiling black holes. 

While the first public use of the term ``black hole'' is generally
attributed to John Wheeler in 1967, as early as 1964 Edwin Saltpeter
proposed that gas accretion onto super-massive black holes provided
the tremendous energy source necessary to power the highly luminous
quasi-stellar objects (quasars) seen in the centers of some galaxies
\cite{saltpeter:64}. Even earlier than that, black holes were
understood to be formal mathematical solutions to Einstein's field
equations \cite{schwarzschild:16}, although considered by many to be
simply mathematical oddities, as opposed to objects that might
actually exist in nature (perhaps most famously, Eddington's stubborn
opposition to the possibility of astrophysical black holes probably
delayed significant progress in their understanding for decades)
\cite{thorne:94}.

In 1969, Lynden-Bell outlined the foundations for black hole accretion
as the basis for quasar power \cite{lynden_bell:69}. The
steady-state thin disks of Shakura and Sunyaev \cite{shakura:73},
along with the relativistic modifications given by Novikov and Thorne
\cite{novikov:73}, are still used as the standard models for accretion
disks today. In the following decade, a combination of theoretical
work and multi-wavelength observations led to a richer understanding
of the wide variety of accretion phenomena in active galactic nuclei
(AGN) \cite{rees:84}. In addition to the well-understood thermal disk
emission predicted by \cite{shakura:73,novikov:73}, numerous non-thermal
radiative processes such as synchrotron and inverse-Compton are
also clearly present in a large fraction of AGN \cite{oda:71,elvis:78}. 

Peters and Mathews \cite{peters:63} derived the leading-order
gravitational wave 
emission from two point masses more than a decade before Thorne and
Braginsky \cite{thorne:76} suggested that one of the most promising
sources for such a GW signal would be the collapse and formation of a
SMBH, or the (near head-on) collision of two such objects
in the center of an active galaxy. In that same paper, Thorne
and Braginsky build on earlier work by Estabrook and Wahlquist
\cite{estabrook:75} and explore the prospects for a space-based
method for direct detection of these GWs via Doppler tracking of
inertial spacecraft. They also attempted to estimate event rates for
these generic bursts, and arrived at quite a broad range of
possibilities, from $\lesssim 0.01$ to $\gtrsim 50$ events per year,
numbers that at least bracket our current best-estimates for SMBH
mergers \cite{sesana:07}. 

However it is not apparent that Thorne and Braginsky
considered the hierarchical merger of galaxies as the driving force
behind these SMBH mergers, a concept that was only just emerging at
the time \cite{ostriker:75,ostriker:77}.
Within the galactic merger context, the seminal paper by Begelman,
Blandford, and Rees (BBR) \cite{begelman:80} outlines the major stages of
the SMBH merger: first the nuclear star clusters merge via dynamical
friction on the galactic dynamical time $t_{\rm gal} \sim 10^8$ yr;
then the SMBHs sink to the center of the new stellar cluster on the
stellar dynamical friction time scale $t_{\rm df} \sim 10^6$ yr; the two SMBHs
form a binary that is initially only loosely bound, and hardens via
scattering with the nuclear stars until the loss cone is depleted;
further hardening is limited by the diffusive replenishing of the loss
cone, until the binary becomes ``hard,'' i.e., the binary's orbital velocity is
comparable to the local stellar orbital velocity, at which point the
evolutionary time scale is $t_{\rm hard} \sim N_{\rm inf} t_{\rm
  df}$, with $N_{\rm inf}$ stars within the influence radius. This is
typically much longer than the Hubble time, 
effectively stalling the binary merger before it can reach the point
where gravitational radiation begins to dominate the evolution. Since
$r_{\rm hard} \sim 1$ pc, and gravitational waves don't take over until
$r_{\rm GW} \sim 0.01$ pc, this loss cone depletion has become known
as the ``final parsec problem'' \cite{merritt:05}. BBR thus propose that
there should be a large cosmological population of stalled SMBH
binaries with separation of order a parsec, and orbital periods of
years to centuries. Yet to date not a single binary system with these
sub-parsec separations has even been unambiguously identified.

In the decades since BBR, numerous
astrophysical mechanisms have been suggested as the solution to the
final parsec problem \cite{merritt:05}. Yet the very fact that so many different
solutions have been proposed and continue to be proposed is indicative
of the prevailing opinion that it is still a real impediment to the
efficient merger of SMBHs following a galaxy merger. However, the
incontrovertible evidence that galaxies regularly undergo minor and
major mergers during their lifetimes, coupled with a distinct lack of
binary SMBH candidates, strongly suggest that nature has found its own
solution to the final parsec problem. Or, as Einstein put it, ``God
does not care about mathematical difficulties; He integrates
empirically.'' 

For incontrovertible evidence of a SMBH binary, nothing can
compare with the direct detection of gravitational waves from space. 
The great irony of gravitational-wave astronomy is that, despite the
fact that the peak GW luminosity generated by black hole mergers
outshines the {\it entire observable universe}, the extremely weak
coupling to matter makes both direct and indirect detection
exceedingly difficult. For GWs with frequencies less than $\sim 1$ Hz, the
leading instrumental concept for nearly 25
years now has been a long-baseline laser interferometer with three
free-falling test masses housed in drag-free spacecraft
\cite{faller:89}. Despite the flurry of recent political and budgetary
constraints that have resulted in a number of alternative, less
capable designs, we
take as our fiducial detector the classic LISA (Laser Interferometer
Space Antenna) baseline design \cite{yellowbook:11}. 

For SMBHs with masses of $10^6
M_\odot$ at a redshift of $z=1$, LISA should be able to identify the
location of the source on the sky within $\sim 10$ deg$^2$ a month
before merger, and better than $\sim 0.1$ deg$^2$ with the
entire waveform, including merger and ringdown
\cite{kocsis:06,lang:06,lang:08,kocsis:08a,lang:09,thorpe:09,mcwilliams:10}.
This should almost certainly be sufficient to identify EM counterparts
with wide-field surveys such as LSST \cite{lsst:09}, WFIRST
\cite{spergel:13}, or WFXT \cite{wfxt:12}. Like the cosmological beacons
of gamma-ray bursts and quasars, merging SMBHs can teach us about
relativity, high-energy astrophysics, radiation hydrodynamics, dark
energy, galaxy formation and evolution, and how they all interact.

A large variety of potential
EM signatures have recently been proposed, almost all of which require
some significant amount of gas in the near vicinity of the merging
black holes \cite{schnittman:11}. Thus we must begin with the question of whether or not
there {\it is} any gas present, and if so, what are its properties. Only
then can we begin to simulate realistic spectra and light curves, and
hope to identify unique observational signatures that will allow us to
distinguish these objects from the myriad of other high-energy
transients throughout the universe.

%--------------------------------------------------------
\section{CIRCUMBINARY DISKS}
\label{disk_theory}
%--------------------------------------------------------

If there is gas present in the vicinity of a SMBH binary, it is likely
in the form of an accretion disk, as least at some point in the
system's history. Disks are omnipresent in the universe for the simple
reason that it is easy to lose energy through dissipative processes,
but much more difficult to lose angular momentum. At larger
separations, before the SMBHs form a bound binary system, massive gas
disks can be quite efficient at bringing the two black holes together
\cite{escala:05,dotti:07}. As these massive gas disks are typically
self-gravitating, their dynamics can be particularly complicated, and
require high-resolution 3D simulations, which will be discussed in
more detail in section \ref{MHD_simulations}.

Here we focus on the properties of non-self-gravitating circumbinary
accretion disks which have traditionally employed the same alpha
prescription for pressure-viscous stress scaling as in \cite{shakura:73}.
Much of the early work on this subject was applied to
protoplanetary disks around binary stars, or stars with massive
planets embedded in their surrounding disks.
The classical work on this subject is Pringle (1991)
\cite{pringle:91}, who considered the evolution of a 1D thin disk with
an additional torque term added to the inner disk. This source of angular
momentum leads to a net outflow of matter, thus giving these systems
their common names of ``excretion'' or ``decretion'' disks. 
Pringle considered two inner boundary conditions: one for the inflow
velocity $v^r(R_{\rm in})\to 0$ and one for the surface density
$\Sigma(R_{\rm in}) \to 0$. For the former case, the torque is applied
at a single radius at the inner edge, leading to a surface density
profile that increases steadily inwards towards $R_{\rm in}$.
In the latter case, the torque is applied over a finite region in the
inner disk, which leads to a relatively large evacuated gap out to
$\gtrsim 6 R_{\rm in}$. In both cases, the angular momentum is
transferred from the binary outwards through the gas disk, leading to
a shrinking of the binary orbit. 

In \cite{artymowicz:91}, SPH simulations were utilized to understand
in better detail the torquing mechanism between the gas and disk. They
find that, in agreement with the linear theory of \cite{goldreich:79},
the vast majority of the binary torque is transmitted to the gas
through the $(l,m)=(1,2)$ outer Lindblad resonance (for more on
resonant excitation of spiral density waves, see \cite{takeuchi:96}).
The resonant interaction
between the gas and eccentric binary ($e=0.1$ for the system in
\cite{artymowicz:91}) pumps energy and angular momentum into the gas,
which gets pulled after the more rapidly rotating interior point
mass. This leads to a nearly evacuated disk inside of $r\approx 2a$,
where $a$ is the binary's semi-major axis. The interaction with the
circumbinary disk not only removes energy and angular momentum from
the binary, but it can also increase its eccentricity, and cause the
binary pericenter to precess on a similar timescale, all of which
could lead to potentially observable effects in GW observations
\cite{armitage:05,roedig:11,roedig:12}.   

In \cite{artymowicz:94,artymowicz:96}, Artymowicz \& Lubow expand upon
\cite{artymowicz:91} and provide a comprehensive study of the effects of
varying the eccentricity, mass ratio, and disk thickness on the
behavior of the circumbinary disk and its interaction with the
binary. Not surprisingly, they find that the disk truncation radius
moves outward with binary eccentricity. Similarly, the mini accretion
disks around each of the stars has an outer truncation radius that
decreases with binary eccentricity. On the other hand, the location of
the inner edge of the circumbinary disk appears to be largely
insensitive to the binary mass ratio \cite{artymowicz:94}. For
relatively thin, cold disks with aspect ratios $H/R\approx 0.03$, the
binary torque is quite effective at preventing accretion, much as in
the decretion disks of Pringle \cite{pringle:91}. In that case, the
gas accretion rate across the inner gap is as much at $10-100\times$
smaller than that seen in a single disk, but the authors acknowledge
that the low resolution of the SPH simulation makes these estimates
inconclusive \cite{artymowicz:94}. 

When increasing the disk thickness
to $H/R\approx 0.1$, the gas has a much easier time jumping the gap
and streaming onto one of the two stars, typically the smaller
one. For $H/R\approx 0.1$,
the gas accretion rate is within a factor of two of the single-disk
case \cite{artymowicz:96}. The accretion rate across the gap is
strongly modulated at the 
binary orbital period, although the accretion onto the individual
masses can be out of phase with each other. The modulated accretion
rate suggests a promising avenue for producing a modulated EM signal
in the pre-merger phase, and the very fact that a significant amount
of gas can in fact cross the gap is important for setting up a
potential prompt signal at the time of merger.

To adequately resolve the spiral density waves in a thin disk, 2D
grid-based calculations are preferable to the inherently noisy and
diffusive SPH methods. Armitage and Natarayan \cite{armitage:02} take
a hybrid approach to the problem, and use a 2D ZEUS \cite{stone:92}
hydrodynamics calculation to normalize the torque term in the 1D
radial structure equation. Unlike \cite{artymowicz:91}, they find
almost no leakage across the gap, even for a moderate
$H/R=0.07$. However, they do identify a new effect that is
particularly important for binary black holes, as opposed to
protoplanetary disks. For a mass ratio of $q\equiv m_2/m_1=0.01$, when
a small accretion disk is formed around the primary, the evolution of
the secondary due to gravitational radiation can shrink the binary on
such short time scales that it plows into the inner accretion disk,
building up gas and increasing the mass accretion rate and thus
luminosity immediately preceding merger \cite{armitage:02}. If robust,
this obviously provides a very promising method for generating bright
EM counterparts to SMBH mergers. However, recent 2D simulations by
\cite{baruteau:12} suggest that the gas in the inner disk could
actually flow across the gap back to the outer disk, like snow flying
over the plow. The reverse of this effect, gas piling up in the outer
disk before leaking into the inner disk, has recently been explored by
\cite{kocsis:12a,kocsis:12b}.

In the context of T Tau stars, \cite{gunther:02,gunther:04} developed
a sophisticated simulation tool that combines a polar grid for the
outer disk with a Cartesian grid around the binary to best resolve the
flow across the gap. They are able to form inner accretion disks
around each star, fed by persistent streams from the circumbinary
disk. As a test, they compare the inner region to an SPH simulation
and find good agreement, but only when the inner disks are
artificially fed by some outer source, itself not adequately resolved
by the SPH calculation \cite{gunther:04}. They also see strong
periodic modulation in the accretion rate, due to a relatively large
binary eccentricity of $e=0.5$. 

\begin{figure}
\begin{center}
\includegraphics[width=0.4\textwidth]{macfadyen_Sigma.epsi}
\includegraphics[width=0.5\textwidth]{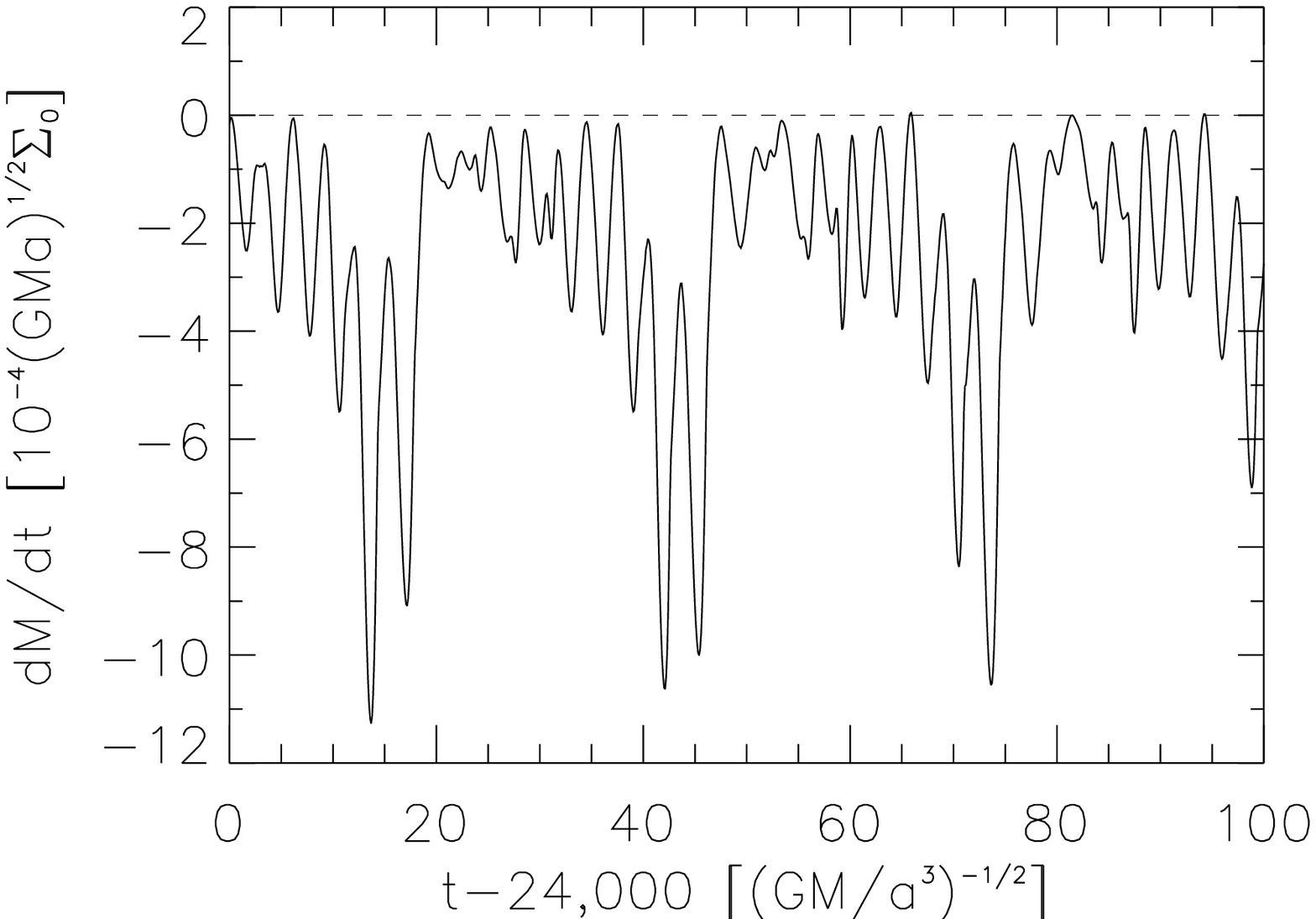}
\caption{\label{fig:macfadyen} ({\it left}) Surface density and spiral
density wave structure of circumbinary disk with equal-mass BHs on a
circular orbit, shown after the disk evolved for 4000 binary
periods. The dimensions of the box are $x=[-5a,5a]$ and
$y=[-5a,5a]$. ({\it right}) Time-dependent accretion rate across the
inner edge of the simulation domain ($r_{\rm in}=a$), normalized by
the initial surface density scale $\Sigma_0$. [reproduced from MacFadyen \&
  Milosavljevic (2008), ApJ {\bf 672}, 83]}
\end{center}
\end{figure}

MacFadyen and Milosavljevic (MM08) \cite{macfadyen:08} also developed a
sophisticated grid-based code including adaptive mesh refinement
to resolve the flows at the inner edge of the circumbinary
disk in the SMBH binary context. However, they excise the inner
region entirely to avoid excessive demands on their resolution around
each black hole so are unable to study the behavior of mini accretion
disks. They also use an alpha prescription for viscosity and 
find qualitatively similar results to the earlier work described
above: a gap with $R_{\rm in} \approx 2a$ due to the $m=2$ outer
Lindblad resonance, spiral density waves in an
eccentric disk, highly variable and periodic accretion, and accretion
across the gap of $\sim 20\%$ that expected for a single BH accretion
disk with the same mass \cite{macfadyen:08}. The disk surface density
as well as the variable accretion rate are shown in Figure
\ref{fig:macfadyen}. Recent work by the same group carried out a
systematic study of the effect of mass ratio and found significant
accretion across the gap for all values of $q=m_2/m_1$ between 0.01
and 1 \cite{dorazio:12}. 

The net result of these calculations seems to be
that circumbinary gas disks are a viable mechanism for driving the
SMBH binary through the final parsec to the GW-driven phase, and
supplying sufficient accretion power to be observable throughout. Thus
it is particularly perplexing that no such systems have been observed
with any degree of certainty. According to simple alpha-disk theory,
there should also be a
point in the GW evolution where the binary separation is shrinking at
such a prodigious rate that the circumbinary disk cannot keep up with
it, and effectively decouples from the binary. At that point, gas
should flow inwards on the relatively slow timescale corresponding to
accretion around a single point mass, and a real gap of evacuated
space might form around the SMBHs, which then merge in a near vacuum
\cite{milos:05}.  

%--------------------------------------------------------
\section{NUMERICAL SIMULATIONS}
\label{simulations}
%--------------------------------------------------------

%--------------------------------------------------------
\subsection{Vacuum numerical relativity}
\label{vacuum}
%--------------------------------------------------------

In the context of EM counterparts, the numerical simulation of two
equal-mass, non-spinning black holes in a vacuum is just about the
simplest problem imaginable. Yet the inherent non-linear behavior of
Einstein's field equations made this a nearly unsolvable Grand
Challenge problem, frustrating generations of relativists from the
3+1 formulation of Arnowitt, Deser, and Misner in 1962
\cite{arnowitt:08}, followed shortly by the first attempt at a
numerical relativity (NR) simulation on a computer in 1964
\cite{hahn:64}, decades of uneven progress, slowed in large part by the
limited computer power of the day (but also by important fundamental
instabilities in the formulation of the field equations), to the ultimate
solution by Pretorius in 2005 \cite{pretorius:05} and subsequent
deluge of papers in 2006 from multiple groups around the world (for a
much more thorough review of this colorful story and the many
technical challenges overcome by its participants, see
\cite{centrella:10}).

Here we will review just a few highlights from the recent NR results
that are most pertinent to our present subject. For the first 50 years
since their original conception,
black holes (and general relativity as a whole) were
largely relegated to mathematicians as a theoretical curiosity with
little possibility of application in astronomy. All this changed in
the late 1960s and early 70s when both stellar-mass and super-massive
black holes were not only observed, but also understood to be critical
energy sources and play a major role in the evolution of galaxies and
stars \cite{thorne:94}. A similar environment was present during the
1990s with regard to binary black holes and gravitational waves. Most
believed in their existence, but after decades of false claims and
broken promises, the prospect of direct detection of GWs seemed
further away than ever. But then in 1999, construction was completed on the two
LIGO observatories, and they began taking science data in 2002. At the
same time, the space-based LISA concept was formalized with the
``Yellow Book,'' a report submitted to ESA in 1996, and together with
NASA, an international science team was formed in 2001. Astrophysics
theory has long been data-driven, but here was a case where
large-scale projects were being proposed and even funded based largely
on theoretical predictions. 

The prospect of real observations and data in turn energized the NR
community and provided new motivation to finally solve the binary BH merger
problem. Long-duration, accurate waveforms are necessary
for both the detection and characterization of gravitational
waves. Generic binary sources are fully described by 17 parameters:
the BH masses (2), spin vectors (6), binary orbital elements (6), sky
position (2), and distance (1). To adequately cover this huge parameter
space requires exceedingly clever algorithms and an efficient method
for calculating waveforms. Fortunately, most NR studies to date
suggest that even the most non-linear phase of the inspiral and merger
process produces a relatively smooth waveform, dominated by the
leading quadrupole mode \cite{centrella:10}. Additionally, in the early inspiral and late
ringdown phases, relatively simple analytic expressions appear to be
quite sufficient in matching the waveforms \cite{pan:11}. Even more encouraging is
the fact that waveforms from different groups using very different
methods agree to a high level of accuracy, thus lending confidence to
their value as a description of the real world \cite{baker:07}.

In addition to the waveforms, another valuable result from these first
merger simulations was the calculation of the mass and spin of the
final black hole, demonstrating that the GWs carried away a full $4\%$
of their initial energy in roughly an orbital time, and leave behind a
moderately spinning black hole with $a/M=0.7$
\cite{baker:06a,campanelli:06}.

After the initial breakthrough with equal-mass, non-spinning black
holes, the remarkably robust ``moving puncture'' method was soon
applied to a wide variety of systems, including unequal masses
\cite{berti:07}, eccentric orbits \cite{hinder:08}, and spinning BHs
\cite{campanelli:06b}. As with test particles around Kerr black holes,
when the spins are aligned with the orbital angular momentum, the BHs
can survive longer before plunging, ultimately producing more GW power
and resulting in a larger final spin. This is another critical result
for astrophysics, as the spin evolution of SMBHs via
mergers and gas accretion episodes is a potentially powerful
diagnostic of galaxy evolution \cite{berti:08}. Perhaps the most
interesting and unexpected result from the NR bonanza was the first accurate
calculation of the gravitational recoil, which will be discussed in
more detail in the following section. 

In addition to the widespread moving puncture method, the NR group at
Cornell/Caltech developed a highly accurate spectral method that is
particularly well-suited for long evolutions \cite{boyle:07}. Because it converges
exponentially with resolution (as opposed to polynomial convergence
for finite-difference methods), the spectral method can generate
waveforms with dozens of GW cycles, accurate to a small fraction of
phase. These long waveforms are particularly useful for matching the
late inspiral to post-Newtonian (PN) equations of motion, the
traditional tool of choice for GW data analysis for LIGO and LISA
(e.g., \cite{cutler:93,apostolatos:94,kidder:95,blanchet:06}). The down 
side of the spectral method has been its relative lack of flexibility,
making it very time consuming to set up simulations of new binary
configurations, particularly with arbitrary spins. If this
problem can be overcome, spectral waveforms will be especially helpful
in guiding the development of more robust semi-analytic tools (e.g., the
effective-one-body approach of Buonanno \cite{buonanno:99}) for calculating the
inspiral, merger, and ringdown of binary BHs with arbitrary initial
conditions. 

The natural application for long, high-accuracy waveforms is as
templates in the matched-filtering approach to GW data analysis. For
LIGO, this is critical to detect most BH mergers, where much of the
in-band power will come from the final stages of inspiral and
merger. The high signal-to-noise expected from SMBHs with LISA
means that most events will probably be detected with high
significance even when using a primitive template library
\cite{flanagan:98,cutler:98}. However,
for {\it parameter estimation}, high-fidelity
waveforms are essential for faithfully reproducing the physical
properties of the source. In particular, for spinning BHs, the
information contained in the precessing waveform can greatly improve
our ability to determine the sky position of the source, and thus
improve our prospects for detecting and characterizing any EM
counterpart \cite{lang:08,thorpe:09,lang:09}.

%--------------------------------------------------------
\subsection{Gravitational recoil}
\label{recoil}
%--------------------------------------------------------

In the general case where there is some asymmetry between the two
black holes (e.g., unequal masses or spins), the GW radiation pattern
will have a complicated multipole structure. The beating between
these different modes leads to a net asymmetry in the momentum flux
from the system, ultimately resulting in a recoil or kick imparted on
the final merged black hole \cite{schnittman:08a}. This effect has long
been anticipated for any GW source
\cite{bonnor:61,peres:62,bekenstein:73}, but the specific value of
the recoil has been notoriously difficult to calculate using
traditional analytic means
\cite{wiseman:92,favata:04,blanchet:05,damour:06}. Because the vast
majority of the recoil is 
generated during the final merger phase, it is a problem uniquely
suited for numerical relativity. Indeed, this was one of the first
results published in 2006, for the merger of two non-spinning BHs with
mass ratio 3:2, giving a kick of $90-100$ km/s \cite{baker:06b}. 

Shortly thereafter, a variety of initial configurations were explored,
covering a range of mass ratios \cite{herrmann:07b,gonzalez:07a}, 
aligned spins \cite{herrmann:07a,koppitz:07}, and
precessing spins \cite{campanelli:07,tichy:07}. Arguably the most exciting result
came with the discovery of the ``superkick'' configuration, where two
equal-mass black holes have equal and opposite spins aligned in the
orbital plane, leading to kicks of $>3000$ km/s
\cite{gonzalez:07b,campanelli:07,tichy:07}. If such a situation 
were realized in nature, the resulting black hole would certainly be
ejected from the host galaxy, leaving behind an empty nuclear host
\cite{merritt:04}. Some of the many other possible ramifications
include offset AGN, displaced star clusters, or unusual accretion
modes. These and other signatures are discussed in detail below in
section \ref{observations}. 

Analogous to the PN waveform matching mentioned above, there has been
a good deal of analytic modeling of the kicks calculated by the NR
simulations
\cite{schnittman:07a,schnittman:08a,boyle:08,racine:09}. Simple
empirical fits to the NR data are particularly useful for
incorporating the effects of recoil into cosmological N-body
simulations that evolve SMBHs along with merging galaxies
\cite{baker:07b,campanelli:07,lousto:09,vanmeter:10b}.
While the astrophysical impacts of large kicks are primarily Newtonian
in nature (even a kick of $v\sim 3000$ km/s is only $1\%$ of the speed
of light), the underlying causes, while only imperfectly understood,
clearly point to strong non-linear gravitational forces at work
\cite{pretorius:07,schnittman:08a,rezzolla:10,jaramillo:12,rezzolla:13}. 

%--------------------------------------------------------
\subsection{Pure electromagnetic fields}
\label{EM_fields}
%--------------------------------------------------------

\begin{figure}
\begin{center}
\includegraphics[width=0.45\textwidth]{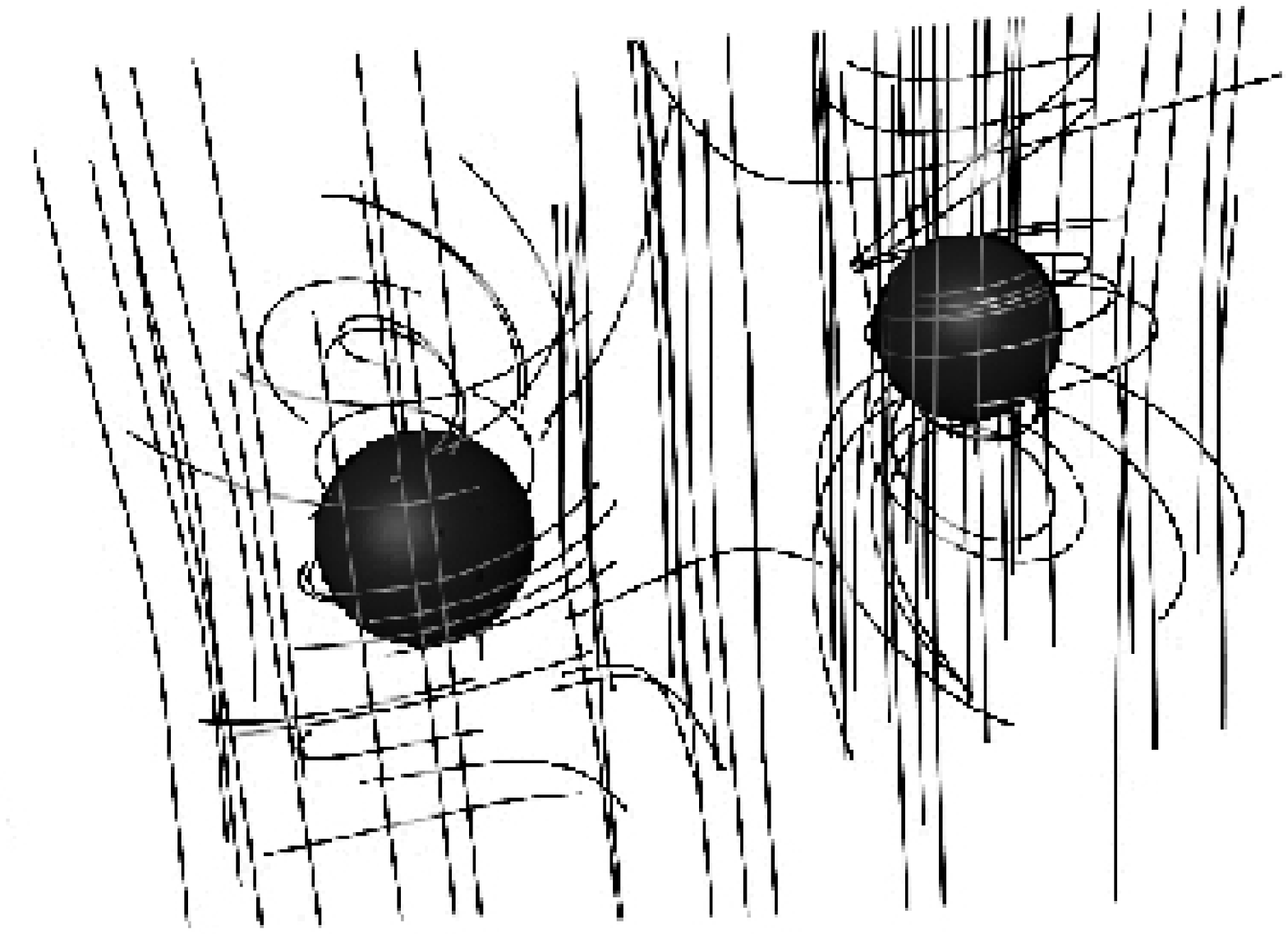}
\includegraphics[width=0.45\textwidth]{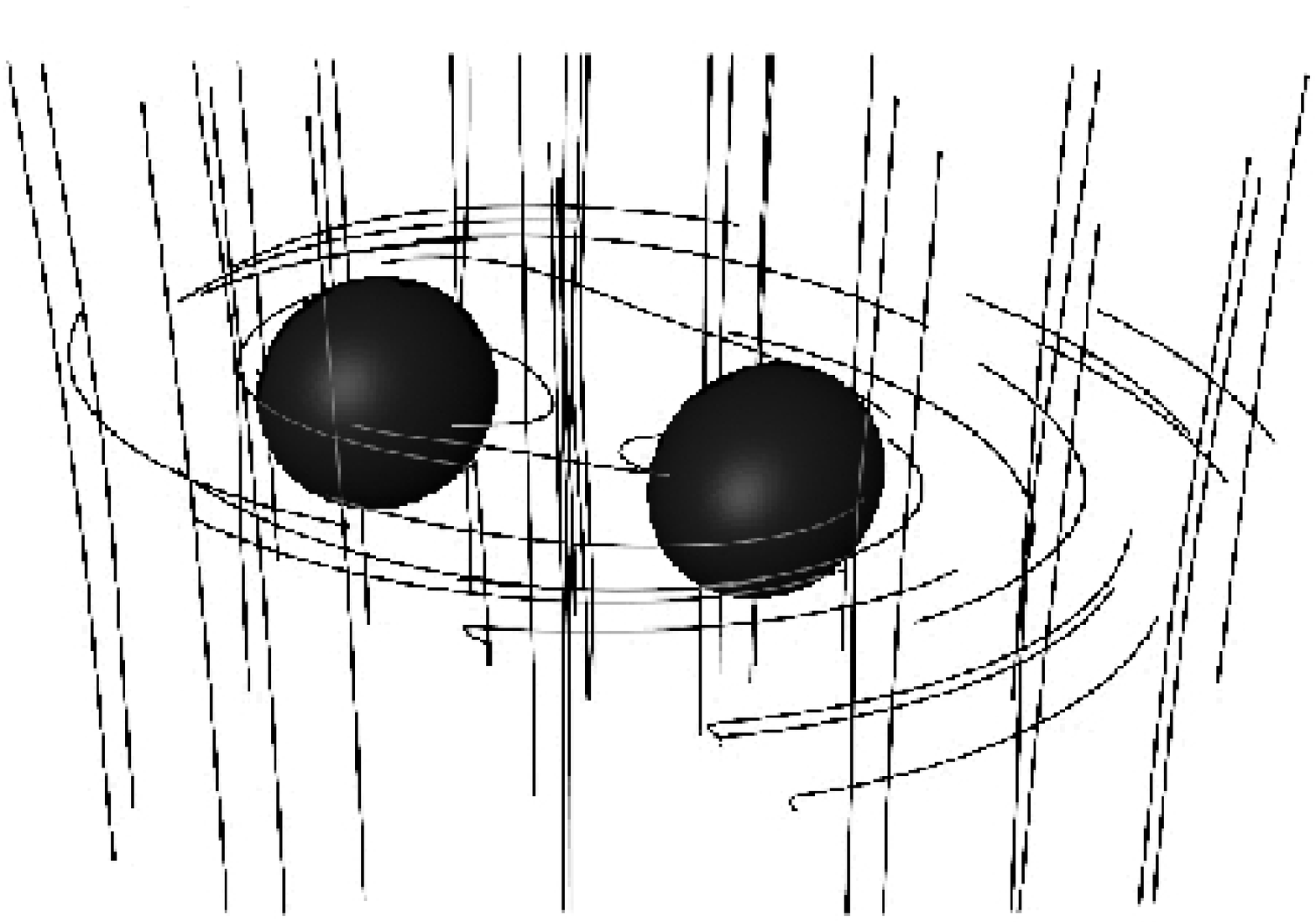}
\caption{\label{fig:palenzuela} Magnetic and electric field
  configurations around binary black hole $40M$ ({\it left}) and $20M$
  ({\it right}) before merger. The electric fields get twisted around
  the black holes, while the magnetic fields remain roughly
  vertical. [reproduced from Palenzuela et al.\ 2009, {\it PRL} {\bf
  103}, 081101]}
\end{center}
\end{figure}

Shortly after the 2006--07 revolution, many groups already began
looking for the next big challenge in numerical relativity. One
logical direction was the inclusion of electromagnetic fields in the
simulations, solving the coupled Einstein-Maxwell equations throughout
a black hole merger. The first to do so was Palenzuela et
al.~\cite{palenzuela:09}, who considered an initial condition with
zero electric field and a uniform magnetic field surrounding an
equal-mass, non-spinning binary a couple orbits before merger. The
subsequent evolution generates E-fields twisted around the two BHs,
while the B-field remains roughly vertical, although it does
experience some amplification (see Fig.\ \ref{fig:palenzuela}). 

The EM power from this system was estimated by integrating the
radial Poynting flux through a spherical shell at large radius. They
found only a modest ($30-40\%$) increase in EM energy, but there was a clear
transient quadrupolar Poynting burst of power coincident with the GW
signal, giving one of the first hints of astrophysical EM counterparts
from NR simulations. This work was followed up by a more thorough
study in \cite{moesta:10,palenzuela:10a}, which showed that the EM
power $L_{\rm EM}$ scaled like
the square of the total BH spin and proportional to $B^2$, as would be
expected for a Poynting flux-powered jet \cite{blandford:77}. 

%--------------------------------------------------------
\subsection{Force-free simulations}
\label{force_free}
%--------------------------------------------------------

In \cite{palenzuela:10b,palenzuela:10c}, Palenzuela and collaborators
extended their vacuum simulations to include force-free
electrodynamics. This is an approximation where a tenuous plasma is
present, and can generate currents and magnetic fields, but carries no
inertia to push those fields around. They found that any moving,
spinning black hole can generate Poynting flux and a
Blandford-Znajek-type jet \cite{blandford:77}. Compared to the vacuum
case, force-free simulations of a merging binary predict significant
amplification of EM power by a factor of $\sim 10 \times$, coincident with
the peak GW power \cite{palenzuela:10c}. For longer simulations run at
higher accuracy, \cite{moesta:12,alic:12} found an even greater $L_{\rm
  EM}$ amplification of $\sim 30 \times$ that of electro-vacuum. 

%--------------------------------------------------------
\subsection{M/HD simulations}
\label{MHD_simulations}
%--------------------------------------------------------

As mentioned above in section \ref{disk_theory}, if there is an
appreciable amount of gas around the binary BH, it is likely in the
form of a circumbinary disk. This configuration has thus been the
subject of most (magneto)hydrodynamical simulations. SPH simulations
of disks that are not aligned with the binary orbit show a warped disk
that can precess as a rigid body, and generally suffer more gas
leakage across the inner gap, modulated at twice the orbital frequency
\cite{larwood:97,ivanov:99,hayasaki:12}. In many cases, accretion
disks can form around the individual BHs \cite{dotti:07,hayasaki:08}.

Massive disks have the
ability to drive the binary towards merger on relatively short time
scales \cite{escala:05,dotti:07,cuadra:09} and also align the BH spins at the same time
\cite{bogdanovic:07} (although see also \cite{lodato:09,lodato:13} for
a counter result). Retrograde disks may be even more efficient at
shrinking the binary \cite{nixon:11} and they may also be quite stable
\cite{nixon:12}. Recent simulations by \cite{roedig:12} show that the 
binary will evolve due not only to torques from the circumbinary disk,
but also from transfer of angular momentum via gas streaming onto the
two black holes. They find that the binary does shrink, and
eccentricity can still be excited, but not necessarily at the rates
predicted by classical theory.

Following merger, the circumbinary disk can also undergo significant
disruption due to the gravitational recoil, as well as the sudden
change in potential energy due to the mass loss from gravitational
waves. These effects lead to caustics forming in the perturbed disk,
in turn leading to shock heating and potentially both prompt and
long-lived EM afterglows
\cite{oneill:09,megevand:09,rossi:10,corrales:10,zanotti:10,
ponce:12,rosotti:12,zanotti:13}. 
Any spin alignment would be
critically important for both the character of the prompt EM
counterpart, as well as the recoil velocity \cite{lousto:12,berti:12}.

Due to computational limitations, it is generally only possible to
include the last few orbits before merger in a full NR
simulation. Since there is no time to allow the system to relax into a
quasi-steady state, the specific choice of initial conditions is
particularly important for these hydrodynamic merger simulations. Some
insight can be gained from Newtonian simulations \cite{shi:12} as well as
semi-analytic models \cite{liu:10,rafikov:12,shapiro:13}.

If the disk decouples from the binary well before merger, the
gas may be quite hot and diffuse around the black holes
\cite{hayasaki:11}. In that case, uniform density diffuse gas may be
appropriate. In merger simulations by
\cite{farris:10,bode:10,bogdanovic:11}, the 
diffuse gas experiences Bondi-type accretion onto each of the SMBHs,
with a bridge of gas connecting the two before merger. Shock
heating of the gas could lead to a strong EM counterpart. As a simple
estimate for the EM signal, \cite{bogdanovic:11} use
bremsstrahlung radiation to predict roughly Eddington luminosity peaking
in the hard X-ray band. 

The first hydrodynamic NR simulations with disk-like initial
conditions were carried out by \cite{farris:11} by allowing the disk
to relax into a quasi-steady state before turning the GR evolution
on. They found disk properties qualitatively similar to classical
Newtonian results, with a low-density gap threaded by accretion
streams at early times, and largely evacuated at late times when the
binary decouples from the disk. Due to the low density and high
temperatures in the gap, they estimate the EM power will be dominated
by synchrotron (peaking in the IR for $M=10^8 M_\odot$), and reach Eddington
luminosity. An analogous calculation was carried out by \cite{bode:12},
yet they find EM luminosity orders of magnitude smaller, perhaps
because they do not relax the initial disk for as long.

Most recently, circumbinary disk simulations have moved from purely
hydrodynamic to magneto-hydrodynamic (MHD), which allows them to
dispense with alpha prescriptions of viscosity and
incorporate the true physical mechanism behind angular momentum in
accretion disks: magnetic stresses and the magneto-rotational
instability \cite{balbus:98}. Newtonian MHD
simulations of circumbinary disks find large-scale $m=1$ modes growing in the outer disk,
modulating the accretion flow across the gap \cite{shi:12}. Similar
modes were seen in \cite{noble:12}, who used a similar procedure as
\cite{farris:11} to construct a quasi-stable state before allowing the
binary to merge. They find that the MHD disk is able to follow the
inspiraling binary to small separations, showing little evidence for
the decoupling predicted by classical disk theory. However, the
simulations of \cite{noble:12} use a hybrid space-time based on PN
theory \cite{gallouin:12} that breaks down close to
merger. Furthermore, while fully relativistic in its MHD treatment,
the individual black holes are excised from the simulation due to
computational limitations, making it difficult to estimate EM
signatures from the inner flow. Farris et al.~\cite{farris:12} have
been able to overcome this issue and put the BHs on the grid with the
MHD fluid. They find that the disk decouples at $a \approx 10M$,
followed by a decrease in luminosity before merger, and then an
increase as the gap fills in and resumes normal accretion, as in
\cite{milos:05}. 

Giacomazzo et al.~\cite{giacomazzo:12} carried out MHD merger
simulations with similar initial conditions to both
\cite{palenzuela:10a} and \cite{bode:10}, with diffuse hot gas
threaded by a uniform vertical magnetic field. Unlike in the
force-free approximation, the inclusion of significant gas leads to a
remarkable amplification of the magnetic field, which is compressed by
the accreting fluid. \cite{giacomazzo:12} found the B-field increased
by of a factor of 100 during merger, corresponding to an increase in
synchrotron power by a factor of $10^4$, which could easily lead to
super-Eddington luminosities from the IR through hard X-ray bands.

The near future promises a self-consistent, integrated picture of
binary BH-disk evolution.
By combining the various methods described above, we can combine
multiple MHD simulations at different scales, using the results from one method
as initial conditions for another, and evolve a circumbinary disk from
the parsec level through merger and beyond.

%--------------------------------------------------------
\subsection{Radiation transport}
\label{radiation}
%--------------------------------------------------------

Even with high resolution and perfect knowledge of the
initial conditions, the value of the GRMHD simulations is limited
by the lack of radiation transport and accurate thermodynamics, which
have only recently been incorporated into local Newtonian simulations of
steady-state accretion disks \cite{hirose:09a,hirose:09b}. Significant
future work 
will be required to incorporate the radiation transport into a fully
relativistic global framework, required not just for accurate modeling
of the dynamics, but also for the prediction of EM signatures that
might be compared directly with observations.

\begin{figure}
\begin{center}
\includegraphics[width=0.65\textwidth]{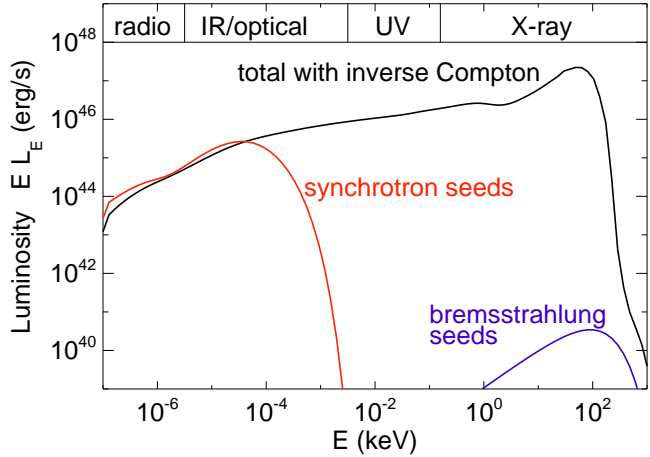}
\caption{\label{fig:pandurata} 
  A preliminary calculation of the broad-band spectrum
  produced by the GRMHD merger of \cite{giacomazzo:12}, sampled near
  the peak of gravitational wave emission. Synchrotron and
  bremsstrahlung seeds from the magnetized plasma are ray-traced with
  {\tt Pandurata} \cite{schnittman:13b}. Inverse-Compton scattering
  off hot electrons in a diffuse corona gives a power-law spectrum
  with cut-off around $kT_e$. The total mass is $10^7 M_\odot$ and the
  gas has $T_e = 100$ keV and optical depth of order unity.}
\end{center}
\end{figure}

Some recent progress has been made by using the relativistic Monte
Carlo ray-tracing code {\tt Pandurata} as a post-processor for MHD
simulations of single accretion disks
\cite{schnittman:13a,schnittman:13b}, reproducing soft and hard X-ray
spectral signatures in agreement with observations of stellar-mass
black holes. Applying the same
ray-tracing approach to the MHD merger simulations of
\cite{giacomazzo:12}, we can generate light curves and broad-band
spectra, ranging from synchrotron emission in the IR up through
inverse-Compton peaking in the X-ray. An example of such a spectrum is
shown in Figure \ref{fig:pandurata}, corresponding to super-Eddington
luminosity at the peak of the
EM and GW emission. Since the simulation in \cite{giacomazzo:12} does
not include a cooling function, we simply estimate the electron
temperature as 100 keV, similar to that seen in typical AGN
coronas. Future work will explore the effects of radiative cooling
within the NR simulations, as well as incorporating the dynamic metric
into the ray-tracing analysis. 

Of course, the ultimate goal will be to directly incorporate radiation
transport as a dynamical force within the GRMHD simulations. Significant
progress has been made recently in developing accurate radiation transport
algorithms in a fully covariant framework
\cite{ohsuga:11,jiang:12,sadowski:13}, and we look forward
to seeing them mature to the point where they can be integrated into
dynamic GRMHD codes. In addition to {\tt Pandurata}, there are a
number of other relativistic
ray-tracing codes (e.g., \cite{dolence:09,shcherbakov:11}), currently
based on the Kerr metric, which may also be adopted to the dynamic
space times of merging black holes.

%--------------------------------------------------------
\section{OBSERVATIONS: PAST, PRESENT, AND FUTURE}
\label{observations}
%--------------------------------------------------------

\begin{figure}
\begin{center}
\includegraphics[width=0.85\textwidth]{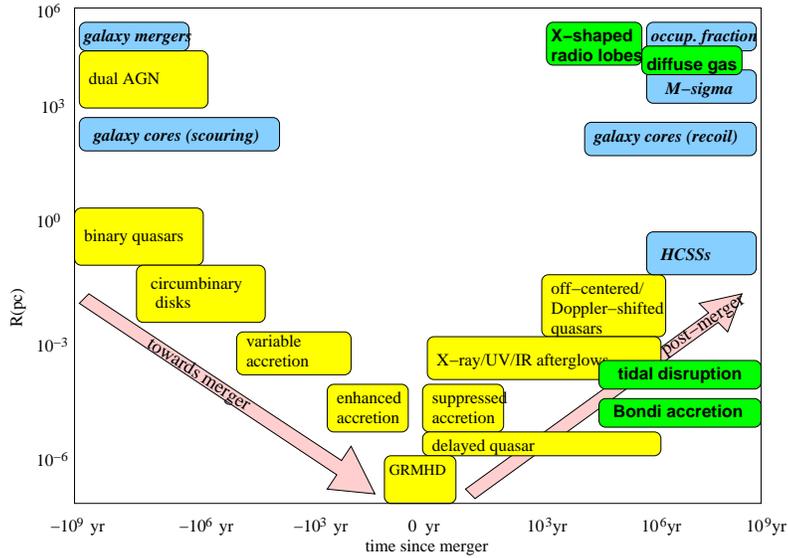}
\caption{\label{source_chart} Selection of potential EM sources,
  sorted by timescale, typical size of emission region, and physical
  mechanism (blue/{\it italic} = stellar; yellow/Times-Roman =
  accretion disk; green/{\bf bold} = diffuse gas/miscellaneous). The
  evolution of the merger proceeds from the upper-left through the
  lower-center, to the upper-right.}
\end{center}
\end{figure}

One way to categorize EM signatures is by the physical mechanism
responsible for the emission: 
stars, hot diffuse gas, or circumbinary/accretion disks. In
Figure \ref{source_chart}, we show the diversity of these sources,
arranged according the spatial and time scales on which they are
likely to occur \cite{schnittman:11}. Over the course of a typical
galaxy merger, we should expect 
the system to evolve from the upper-left to the lower-center to the
upper-right regions of the chart. Sampling over the entire observable
universe, the number of objects detected in each source class should
be proportional to the product of the lifetime and observable flux of
that object. 

Note that most of these effects are fundamentally Newtonian,
and many are only indirect evidence of SMBH mergers, as opposed to the
prompt EM signatures described above. Yet they are also important in
understanding the complete history of binary BHs, as they are crucial
for estimating the number of sources one might expect at each stage in
a black hole's evolution. If, for example, we predict a large number
of bright binary quasars with separations around $0.1$ pc, and find
no evidence for them in any wide-field surveys (as has been the case
so far, with limited depth and temporal coverage), we would be forced to
revise our theoretical models. But if the same rate calculations
accurately predict the number of dual AGN with separations of $\sim
1-10$ kpc, and GW or prompt EM detections are able to confirm the
number of actual mergers, then we might infer the lack of binary quasars is
due to a lack of observability, as opposed to a lack of existence. 

The long-term goal in observing EM signatures will be to eventually
fill out a plot like that of Figure \ref{source_chart}, determining
event rates for each source class, and checking to make sure we can
construct a consistent picture of SMBH-galaxy co-evolution. This is
indeed an ambitious goal, but one that has met with reasonable success
in other fields, such as stellar evolution or even the fossil record
of life on Earth. 

\subsection{Stellar Signatures}

On the largest scales, we have strong circumstantial evidence of
supermassive BH mergers at the centers of merging galaxies. From large
optical surveys of interacting galaxies out to redshifts of $z \sim
1$, we can infer that $5-10\%$ of massive galaxies
are merging at any given time, and the majority of galaxies with
$M_{\rm gal} \gtrsim 10^{10} M_\odot$ have experienced a major merger
in the past 3 Gyr \cite{bell:06,mcintosh:08,deravel:09,bridge:10},
with even higher merger
rates at redshifts $z\sim 1-3$ \cite{conselice:03}. At the same time,
high-resolution observations of nearby galactic nuclei find that every
large galaxy hosts a SMBH in its center \cite{kormendy:95}. Yet we see
a remarkably small number of dual AGN \cite{komossa:03,comerford:09},
and only one known source with an actual binary system where the BHs
are gravitationally bound to each other
\cite{rodriguez:06,rodriguez:09}. Taken
together, these observations strongly suggest that when galaxies
merge, the merger of their central SMBHs inevitably follows, and
likely occurs on a relatively short time scale, which would explain
the apparent scarcity of binary BHs (although recent estimates by
\cite{hayasaki:10} predict as many as $10\%$ of AGNs with $M\sim 10^7
M_\odot$ might be in close binaries with $a\sim 0.01$ pc). The famous
``M-sigma'' 
relationship between the SMBH mass and the velocity dispersion of the
surrounding bulge also points to a merger-driven history over a wide
range of BH masses and galaxy types \cite{gultekin:09}.

There is additional indirect evidence for SMBH mergers in the stellar
distributions of galactic nuclei, with many elliptical galaxies
showing light deficits (cores), which correlate strongly with the
central BH mass \cite{kormendy:09}. The cores suggest a 
history of binary BHs that scour out the nuclear stars via three-body
scattering
\cite{milosavljevic:01,milosavljevic:02,merritt:07}, or even
post-merger relaxation of recoiling BHs
\cite{merritt:04,boylan-kolchin:04,gualandris:08,guedes:09}. 

While essentially all massive nearby galaxies appear to
host central SMBHs, it is quite possible that this is not the case
at larger redshifts and smaller masses, where major mergers could lead
to the complete ejection of the resulting black hole via large
recoils. By measuring the occupation fraction of
SMBHs in distant galaxies, one could infer merger rates and the
distribution of kick velocities \cite{schnittman:07a,volonteri:07,
  schnittman:07b,volonteri:08a,volonteri:10}. The occupation fraction
will of course also affect the LISA event rates, especially at high
redshift \cite{sesana:07}. 

Another indirect signature of BH mergers comes from the population of
stars that remain bound to a recoiling black hole that gets ejected
from a galactic nucleus \cite{komossa:08a,merritt:09,oleary:09}. These
stellar systems will appear similar to
globular clusters, yet with smaller spatial extent and much larger
velocity dispersions, as the potential is completely dominated by the
central SMBH. With multi-object spectrometers on large ground-based telescopes,
searching for these stellar clusters in the Milky Way halo or nearby
galaxy clusters ($d \lesssim 40$ Mpc) is technically realistic
in the immediate future.

\subsection{Gas Signatures: Accretion Disks}

As discussed above in section \ref{disk_theory}, circumbinary disks
will likely have a low-density gap within $r\approx 2a$, although may still
be able to maintain significant gas accretion across this gap, even
forming individual accretion disks around each black hole. The most
sophisticated GRMHD simulations suggest that this accretion can be
maintained even as the binary is rapidly shrinking due to
gravitational radiation \cite{noble:12}. If the inner disks can
survive long enough, the final inspiral may lead to a rapid enhancement of
accretion power as the fossil gas is plowed into the central black
hole shortly before merger \cite{armitage:02, chang:10}. For small
values of $q$, a narrow gap could form in the inner disk, changing the
AGN spectra in a potentially observable
way \cite{gultekin:12,mckernan:13}.

Regardless of {\it how} the gas reaches the central BH region, the
simulations described above in section \ref{simulations} all seem to
agree that even a modest amount of magnetized gas can lead to a strong
EM signature. 
If the primary energy source for heating the gas is gravitational \cite{vanmeter:10},
then typical efficiencies will be on the order of $\sim 1-10$\%,
comparable to that expected for standard accretion in
AGN, although the much shorter timescales could easily lead to
super-Eddington transients, depending on the optical depth and cooling
mechanisms of the gas\cite{krolik:10}. 

However, if the merging BHs are able to generate strong magnetic
fields \cite{palenzuela:09, moesta:10, palenzuela:10b,giacomazzo:12},
then hot electrons could easily generate strong synchrotron flux, or
highly relativistic jets may be launched along the resulting BH spin axis,
converting matter to energy with a Lorentz boost factor of $\Gamma \gg 1$. 
Even with purely hydrodynamic heating,
particularly bright and long-lasting afterglows may be produced
in the case of large recoil velocities, which effectively can
disrupt the entire disk, leading to strong shocks and dissipation
\cite{lippai:08, shields:08, schnittman:08b, megevand:09, rossi:10,
  anderson:10, corrales:10, tanaka:10a, zanotti:10}. 
Long-lived afterglows could be discovered in existing
multi-wavelength surveys, but successfully identifying them as merger
remnants as opposed to obscured AGN or other bright unresolved sources
would require improved pipeline analysis of literally millions of
point sources, as well as extensive follow-up observations
\cite{schnittman:08b}. 

For many of these large-kick systems,
we may observe quasar activity for millions of years after,
with the source displaced from the galactic center, either spatially
\cite{kapoor:76, loeb:07,
  volonteri:08b,civano:10,dottori:10,jonker:10} or spectroscopically
\cite{bonning:07, komossa:08c,boroson:09,robinson:10}. However, large offsets between the
redshifts of quasar emission lines and their host galaxies have also
been interpreted as evidence of pre-merger binary BHs
\cite{bogdanovic:09,dotti:09,tang:09,dotti:10b} or due to the large relative
velocities in merging galaxies \cite{heckman:09,shields:09a,vivek:09,decarli:10}, or
``simply'' extreme examples of the class of double-peaked emitters,
where the line offsets are generally attributed to the disk
\cite{gaskell:88,eracleous:97,shields:09b,chornock:10,gaskell:10}. An indirect signature for kicked BHs
could potentially show up in the statistical properties of active
galaxies, in particular in the relative distribution of different
classes of AGN in the ``unified model'' paradigm
\cite{komossa:08b,blecha:11}. 

For systems that open up a 
gap in the circumbinary disk, another EM signature may take the form of a
quasar suddenly turning on as the gas refills the gap, months to years
after the BH merger \cite{milos:05, shapiro:10, tanaka:10b}. But
again, these sources would be difficult to distinguish from normal AGN
variability without known GW counterparts. Some limited searches for
this type of variability have recently been carried out in the X-ray
band \cite{kanner:13},
but for large systematic searches, we will need targeted time-domain
wide-field surveys like PTF, Pan-STARRS, and eventually LSST. One of
the most valuable scientific products from these time-domain surveys
will be a better understanding of what is the range of variability for
normal AGN, which will help us distinguish when an EM signal is most
likely due to a binary \cite{tanaka:13}. 

In addition to the many potential prompt and afterglow signals from
merging BHs, there has also been a significant amount of
theoretical and observational work focusing on the early precursors of
mergers. Following the evolutionary trail in
Figure 1, we see that shortly after a galaxy merges, dual AGN may
form with typical separations of a few kpc
\cite{komossa:03,comerford:09}, sinking to the center of the merged
galaxy on a relatively short timescale ($\lesssim$ 1 Gyr) due to
dynamical friction \cite{begelman:80}. The galaxy merger process is
also expected to funnel a great deal of gas to the galactic center, in
turn triggering quasar activity
\cite{hernquist:89,kauffmann:00,hopkins:08,green:10}. At separations
of $\sim 1$ pc, the BH binary (now ``hardened'' into a gravitationally
bound system) could stall, having depleted its loss cone of stellar
scattering and not yet reached the point of gravitational radiation
losses \cite{milosavljevic:03}. Gas dynamical drag from
massive disks ($M_{\rm disk} \gg M_{\rm BH}$) leads to a prompt
inspiral ($\sim 1-10$ Myr), in most cases able to reach sub-parsec
separations, depending on the resolution of the simulation
\cite{escala:04,kazantzidis:05,escala:05,dotti:07,
cuadra:09,dotti:09b,dotti:10a}.

At this point, a proper binary quasar is formed, with an orbital period
of months to decades, which could be identified by periodic accretion
\cite{macfadyen:08, hayasaki:08, haiman:09a, haiman:09b}, density
waves in the disk \cite{hayasaki:09}, or periodic red-shifted broad emission
lines \cite{bogdanovic:08,shen:09,loeb:10,montuori:11}. If these
binary AGN systems do in fact exist,
spectroscopic surveys should be able to identify many candidates,
which may then be confirmed or ruled out with subsequent
observations over relatively short timescales ($\sim 1-10$ yrs), as
the line-of-site velocities to the BHs changes by an observable
degree. This approach has been attempted with various initial
spectroscopic surveys, but
as yet, no objects have been confirmed to be binaries by multi-year
spectroscopic monitoring
\cite{boroson:09,lauer:09,chornock:10,eracleous:12}.

\subsection{Gas Signatures: Diffuse Gas; ``Other''}
In addition to the many disk-related signatures, there are also a
number of potential EM counterparts that are caused by the accretion
of diffuse gas in the galaxy. For the Poynting flux generated by the
simulations of section \ref{simulations}, transient bursts or modulated
jets might be detected in all-sky radio surveys \cite{kaplan:11,oshaughnessy:11}.
For BHs that get significant kicks at
the time of merger, we expect to see occasional episodes of Bondi
accretion as the BH oscillates through the gravitational potential
of the galaxy over millions of years, as well as off-center AGN
activity \cite{blecha:08, fujita:09,guedes:10,sijacki:10}. On larger
spatial scales, the recoiling BH could also produce trails of
over-density in the hot interstellar gas of elliptical galaxies
\cite{devecchi:09}.
Also on kpc--Mpc scales, X-shaped radio jets have been seen in a
number of galaxies, which could possibly be due to the merger and
subsequent spin-flip of the central BHs \cite{merritt:02}.

Another potential source of EM counterparts comes not from diffuse
gas, or accretion disks, but the occasional capture and tidal
disruption of normal stars by the merging BHs. These tidal disruption
events (TDEs),
which also occurs in ``normal'' galaxies \cite{rees:88,komossa:99,halpern:04},
may be particularly easy to identify in off-center BHs following a
large recoil \cite{komossa:08a}. TDE rates may be strongly
increased prior to the merger 
\cite{chen:09,stone:10,seto:10,schnittman:10,chen:11,wegg:11}, but the actual
disruption signal may be truncated by the pre-merger binary
\cite{liu:09}, and post-merger recoil may also reduce the rates
\cite{li:12}. These TDE events are likely to be seen by the dozen in 
coming years with Pan-STARRS and LSST \cite{gezari:09}. In addition to
the tidal disruption scenario, in 
\cite{schnittman:10} we showed how gas or stars trapped at the stable
Lagrange points in a BH binary could evolve during inspiral and
eventually lead to enhanced star formation, ejected hyper-velocity
stars, highly-shifted narrow emission lines, and short bursts of
super-Eddington accretion coincident with the BH merger. 

A completely different type of EM counterpart can be seen with pulsar
timing arrays (PTAs). In this technique, small time delays ($\lesssim 10$ ns) in
the arrival of pulses from
millisecond radio pulsars would be direct evidence of extremely
low-frequency (nano-Hertz) gravitational waves from massive ($\gtrsim
10^8 M_\odot$) BH binaries
\cite{jenet:06,sesana:08,sesana:09,jenet:09,seto:09,
pshirkov:10,vanhaasteren:10,sesana:10}. 
By cross-correlating the signals from multiple pulsars around the sky,
we can effectively make use of a GW detector the size of the entire
galaxy. For now, one of the main impediments to GW astronomy with
pulsar timing is the relatively small number of known, stable
millisecond radio pulsars. Current surveys are working to increase
this number and the uniformity of their distribution on the sky
\cite{lee:13}. 

Even conservative estimates suggest that PTAs are probably only about
ten years away from a positive detection of the GW stochastic
background signal from the ensemble of SMBH binaries throughout the
universe \cite{sesana:13}. The probability of resolving an individual source is
significantly smaller, but if it were detected, would be close enough
($z \lesssim 1$) to allow for extensive EM follow-up, unlike many of
the expected LISA sources at $z \gtrsim 5$. Also, unlike LISA sources,
PTA sources would be at an earlier stage in their inspiral and thus be
much longer lived, allowing for even more extensive study. A
sufficiently large sample of such sources would even allow us to test
whether they are evolving due to GW emission or gas-driven migration
\cite{kocsis:11,tanaka:12,sesana:12} (a test that might also be done
with LISA with only a single source with sufficient signal-to-noise
\cite{yunes:11}). 

\section{CONCLUSION}

Black holes are fascinating objects. They push our intuition to the
limits, and never cease to amaze us with their extreme behavior. For a
high-energy theoretical astrophysicist, 
the only thing more exciting than a real astrophysical
black hole is {\it two} black holes, destroying everything in their
path as they spiral together towards the point of no return. Thus one
can easily imagine the frustration that stems from our lack of ability
to actually see such an event, despite the fact that it outshines the
entire observable universe. And the path forwards does not appear to
be a quick one, at least not for gravitational-wave astronomy. 

One important step along this path is the engagement of the broader
(EM) astronomy community. Direct detection of gravitational waves will
not merely be a confirmation of a century-old theory---one more
feather in Einstein's Indian chief
head-dress---but the opening of a window through which we can observe the
entire universe at once, eagerly listening for the next thing to go
bang in the night. And when it does, all our EM eyes can swing over to
watch the fireworks go off. With a tool as powerful as coordinated
GW/EM observations, we will be able to answer many of the
outstanding questions in astrophysics: 

How were the first black holes formed?
Where did the first quasars come from? What is the galaxy merger rate as a
function of galaxy mass, mass ratio, gas fraction, cluster
environment, and redshift? What is the mass function and spin
distribution of the central BHs in these merging (and non-merging)
galaxies? What is the central environment around the BHs, prior to
merger: What is the quantity and quality (temperature, density,
composition) of gas? What is the stellar distribution (age, mass
function, metallicity)? What are the properties of the circumbinary
disk? What is the time delay between galaxy merger and BH merger? 

These are just a few of the mysteries that will be solved with the
routine detection and characterization of SMBH mergers, may we witness
them speedily in our days!

\vspace{0.5cm}

We acknowledge helpful conversations with John Baker, Manuela
Campanelli, Bruno Giacomazzo, Bernard Kelly, Julian Krolik, Scott
Noble, and Cole Miller.

\end{document}